\documentclass[prd,aps,showpacs,tightenlines]{revtex4}  
\usepackage{mathrsfs}
\usepackage{amsmath}
\usepackage{amssymb}
\usepackage{epsfig}
\usepackage{graphicx}
\usepackage{booktabs}
\usepackage{multirow}
\usepackage{subfigure}

\begin{document}
\newcommand{\psl}{ p \hspace{-1.8truemm}/ }
\newcommand{\nsl}{ n \hspace{-2.2truemm}/ }
\newcommand{\vsl}{ v \hspace{-2.2truemm}/ }
\newcommand{\epsl}{\epsilon \hspace{-1.8truemm}/\,  }

\title{Probing  the $P$-wave charmonium decays of $B_c$ meson}
\author{Zhou Rui$^1$}\email{jindui1127@126.com}
\affiliation{$^1$College of Sciences, North China University of Science and Technology,
                          Tangshan 063009,  China}
\date{\today}

\begin{abstract}

Motivated by the large number of $B_c$ meson decay modes observed recently by  several detectors at the LHC,
we present a detailed analysis of the $B_c$ meson decaying to the $P$-wave charmonium states and a light pseudoscalar
 ($P$) or vector ($V$) meson within the framework of perturbative QCD factorization.
 The $P$-wave charmonium distribution amplitudes are extracted from the $n=2,l=1$ Schr$\ddot{o}$dinger states for a Coulomb potential,
 which can be taken  as the universal nonperturbative objects to analyze the hard exclusive processes with $P$-wave charmonium production.
It is found that these decays have  large branching ratios of the order of $10^{-5}\sim 10^{-2}$,
 which seem to be in the reach of future experiments.
  We also provide predictions for the polarization fractions and relative phases  of $B_c\rightarrow (\chi_{c1},\chi_{c2},h_c)V$ decays.
It is expected  that  the longitudinal polarization amplitudes  dominate the branching ratios according to the quark helicity analysis,
and the magnitudes and phases of parallel polarization amplitude are approximately equal to the perpendicular ones.
The obtained results are compared with available experimental data, our previous studies, and numbers from other approaches.
\end{abstract}

\pacs{13.25.Hw, 12.38.Bx, 14.40.Nd }


\maketitle

\section{Introduction}

\begin{table}
\caption{The properties of $P$-wave charmonium states \cite{pdg2016}.}
\label{tab:properties}
\begin{tabular}[t]{p{1.5cm}p{1.5cm}p{1.5cm}p{2.5cm}p{2.5cm}}
\hline\hline
Mesons    & $n{}^{2s+1}L_J$ &$J^{PC}$ & Mass (MeV) &Width (MeV)\\
$\chi_{c0}$ & $1{}^{3}P_0$ & $0^{++}$  &$3414.75\pm 0.31$  & $10.5\pm 0.6$ \\
$\chi_{c1}$ & $1{}^{3}P_1$ & $1^{++}$  &$3510.66\pm 0.07$ & $0.84 \pm 0.04$\\
$\chi_{c2}$ & $1{}^{3}P_2$ & $2^{++}$  &$3556.20\pm 0.11$ & $1.93 \pm 0.11$ \\
$h_{c}$     &$1{}^{1}P_1$ &  $1^{+-}$ &$3525.38\pm 0.11$ & $0.7 \pm 0.4$\\
\hline\hline
\end{tabular}
\end{table}
In the quark model, $P$-wave charmonium states are expected as the orbital excitation of the $c\bar{c}$  assignments
with the orbital angular momentum $L=1$.
 Since the charm-anticharm quarks pair can be in the spin singlet or spin triplet states,
in terms of the spectroscopic notation ${}^{2s+1}L_J$, there are four types of $P$-wave charmonium states, namely, $\chi_{c0}({}^{3}P_0)$,
$\chi_{c1}({}^{3}P_1)$, $\chi_{c2}({}^{3}P_2)$, and $h_{c}({}^{1}P_1)$.
The current experimental knowledge of these $P$-wave charmonium states is summarized in Table \ref{tab:properties} \cite{pdg2016}.
 Experimentally the productions of these $P$-wave charmonium states have been seen in the   hadronic $B$ decays:
 $B\rightarrow \chi_{c1} \pi$ \cite{prd74051103,prd78091104},
$B\rightarrow \chi_{c0} K^{*}$ \cite{prl88031802,prd78091101}, $B\rightarrow \chi_{c1,c2} K^{(*)}$ \cite{prl89011803,prl94141801,prl102132001,plb643155,prl107091803},
$B\rightarrow h_{c} K^{(*)}$ \cite{prd74012007,prd78012006},  $B_s\rightarrow \chi_{c1} \phi$ \cite{plb874663}, 
$B\rightarrow \chi_{c1,c2}K\pi$ \cite{prd85052003,prd93052016}, $B\rightarrow \chi_{c1,c2}\pi\pi K$ \cite{prd93052016},
and  in $\Lambda_b^0$ baryon decay: $\Lambda_b^0\rightarrow \chi_{c1,c2}pK^-$ \cite{prl119062001}.
As for   hadronic $B_c$ decays,  the first evidence of $B^+_c\rightarrow \chi_{c0}(\rightarrow K^+K^-)\pi^+ $  \cite{prd94091102}
is reported with a significance of 4.0 standard deviations by  the LHCb experiment.
The measured product of the ratio of cross sections and branching fraction is
\begin{eqnarray}
\frac{\sigma_{B_c^+}}{\sigma_{B^+}}\times \mathcal {B}(B^+_c\rightarrow \chi_{c0}\pi^+)
=(9.8^{+3.4}_{-3.0}(\text{stat})\pm 0.8(\text{syst}))\times 10^{-6},
\end{eqnarray}
where  $\sigma_{B_c^+}(\sigma_{B^+})$ is the production cross sections for $B_c(B)$ meson.

As is well known, the pseudoscalar $B_c$ is composed of two  heavy-flavored quarks
and thus represents a unique laboratory  to study heavy-quark dynamics and $CP$ violation.
Since each of the two heavy quarks can decay with the other as a spectator,
the $B_c$ meson has rich decay channels, and offers a promising  opportunity
to study nonleptonic weak decays of heavy mesons, to test the standard model (SM), and even to reveal any new physics beyond SM.
Decays of $B_c$ mesons to the final states including a
charmonium meson are of special interest. First, these decay
modes   provide a sensitive laboratory for studying strong
interaction   effects in a heavy meson system.
Second, those decays involve two energy scales, the bottom quark mass $m_b$ and charm quark mass $m_c$.
The higher order corrections within the framework of quantum chromodynamics (QCD), described by the expansion  of $m_c/m_b$ rather than $\Lambda_{QCD}/m_b$ with $\Lambda_{QCD}$ is the QCD scale, may be relatively large, and therefore are more subtle in theoretical studies.
 Third, one can search for charmonium and charmoniumlike exotic states in one of
the intermediate final states such as $\chi_{cJ}\pi$ and $\chi_{cJ}\pi\pi$ ($J=0,1,2$), which may be  important to understand the detailed
dynamics of the multibody $B_c$ decay modes.
Besides, such $B_c$ decays provide a direct probe of charmonium properties
by reconstructing the charmonium state from its decay to some known final state.

Phenomenologically the $B_c$ meson  decays  into  various charmonium states  have been widely studied
in the literature. Earlier, a lot of work has been done in the semileptonic
and nonleptonic \cite{prd493399,prd564133,prd62014019,prd61034012,npb585353,prd68094020,
jhep06015,prd86094028,prd88094014,prd89034008,prd89017501,scp58071001} decays of the meson $B_c$ to $S$-wave charmonium
mesons. Also, the $P$-wave charmonium  decays of $B_c$ meson
have been considered previously by other authors
\cite{prd82034019,prd74074008,prd71094006,prd73054024,prd77054003,prd65014017,jpg28595,jpg39015009,prd87034004,prd79114018}.
Furthermore, some semileptonic and nonleptonic decays of $B_c$  into the $D$-wave charmonium states
have been analyzed  in the framework of the instantaneous Bethe-Salpeter method \cite{epjc76454}.
More recently, the exclusive decays of the $B_c$ meson into $P$-wave orbitally excited charmonium and
a light meson have been investigated using the nonrelativistic QCD effective theory \cite{171007011},
 where the next-to-leading order relativistic corrections to the corresponding form factors are considered.

As a successive work of \cite{rui1,rui2},
in the present work we will focus on the $B_c$ decays
involving a $P$-wave charmonium state  and a light pseudoscalar or vector
meson  in the final states employing the Perturbative QCD (PQCD) approach based on the $k_T$ factorization theorem.
Similar to the case of $S$-wave charmonium states \cite{plb612215},
the $P$-wave charmonium  distribution amplitudes (DAs)
  can also be expressed  as  an   associated factor,  extracted from the $P$-wave Schr$\ddot{o}$dinger states for a Coulomb potential,
   multiply by the asymptotic models of the corresponding twists for light mesons.
   With the help of  $P$-wave DAs, we can make quantitative predictions here, and
 provide a ready reference to existing and forthcoming experiments.

The rest of this article is organised in the following way.
In Sec. \ref{sec:framework}, the Hamiltonian and kinematics, and
the $P$-wave charmonium DAs  are shown in cases of   scalar, axial-vector, and tensor states. Then
the calculations of these decay amplitudes in the PQCD framework are briefly reviewed.
In Sec. \ref{sec:results}, the adopted parameters, numerical results and discussion are given in detail.
Finally, the conclusions are given in Sec. \ref{sec:sum}.
The evaluation of the $P$-wave charmonium distribution
amplitudes is relegated to the Appendix.

\section{Formalism }\label{sec:framework}
 \subsection{HAMILTONIAN AND KINEMATICS }\label{sec:handk}

 The effective Hamiltonian describing the $B_c$ nonleptonic decays into charmonium and a light pseudoscalar or vector meson
is given by \cite{rmp681125}
 \begin{eqnarray}
\mathcal {H}_{eff}=\frac{G_F}{\sqrt{2}}V^*_{cb}V_{uq}[C_1(\mu)O_1(\mu)+C_2(\mu)O_2(\mu)],
\end{eqnarray}
 where $q=s,d$ stands for a down type light quark. $G_F$ is the Fermi constant. $V^*_{cb}$ and $V_{uq}$ are the Cabibbo-Kobayashi-Maskawa
(CKM) matrix elements.
 $C_{1,2}(\mu)$ are the perturbatively calculable Wilson coefficients,
 which encode the short-distance effects above the renormalization scale $\mu$,
 while  $O_{1,2}(\mu)$ are the corresponding  local four-quark operators,
whose expressions  read as
 \begin{eqnarray}
O_1(\mu)&=&\bar{b}_{\alpha}\gamma^{\nu}(1-\gamma_5)c_{\beta}\otimes \bar{u}_{\beta}\gamma_{\nu}(1-\gamma_5)q_{\alpha},\nonumber\\
O_2(\mu)&=&\bar{b}_{\alpha}\gamma^{\nu}(1-\gamma_5)c_{\alpha}\otimes \bar{u}_{\beta}\gamma_{\nu}(1-\gamma_5)q_{\beta},
\end{eqnarray}
where $\alpha$ and $\beta$ are color indices and the summation convention
over repeated indices is understood.
Since the  Hamiltonian involves  four different flavor quarks, it means that these decays are uncontaminated by
the contributions from the penguin operators, and thus the direct $CP$ asymmetries are absent naturally.

\begin{figure}[tbp]
\centerline{\epsfxsize=7cm \epsffile{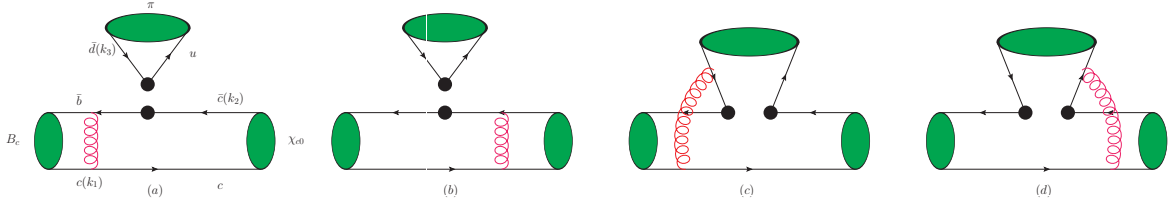}}
\vspace{-6cm}
\caption{The typical leading-order Feynman diagrams for the decay  $B_c \to \chi _{c0} \pi$.
(a,b) The factorizable  diagrams,  and (c,d) the nonfactorizable diagrams.}
\label{fig:femy}
\end{figure}

The calculation is carried out in the rest frame of $B_c$ meson, the  $B_c$ meson momentum $P_1$,
the recoiled charmonium meson momentum $P_2$, and the ejected light meson momentum $P_3$ 
are defined in the light cone coordinates as
\begin{eqnarray}
 P_1&=&\frac{M}{\sqrt{2}}(1,1,\textbf{0}_{\rm T}),\quad P_2=\frac{M}{\sqrt{2}}(1-r_3^2,r_2^2,\textbf{0}_{\rm T}),\quad  P_3=\frac{M}{\sqrt{2}}(r_3^2,1-r_2^2,\textbf{0}_{\rm T}),
\end{eqnarray}
with the mass ratio $r_{2,3}=m_{2,3}/M$ and $M$ ($m_{2}$) is  the $B_c$ (charmonium) meson mass, while
$m_3$ is the (chiral) mass of the  (pseudoscalar) vector meson.
The momentum of the valence quarks $k_{1,2,3}$, whose notation is displayed in Fig. \ref{fig:femy}, is parametrized as
\begin{eqnarray}
 k_1&=&x_1P_1+\textbf{k}_{\rm 1T},\quad k_2=x_2P_2+\textbf{k}_{\rm 2T},\quad  k_3=x_3P_3+\textbf{k}_{\rm 3T},
\end{eqnarray}
where $k_{iT}$ , $x_i$ represent the transverse momentum and longitudinal momentum fraction of the quark/anti-quark inside the meson.
When the final states contain a axial-vector charmonium  and a vector meson, the longitudinal
polarization vectors $\epsilon_L$ and transverse polarization vectors $\epsilon_T$ can be defined as
\begin{eqnarray}\label{eq:polar}
\epsilon_{2L}&=&\frac{1}{\sqrt{2(1-r_3^2)}r_2}(1-r_3^2,-r_2^2,\textbf{0}_{\rm T}), \quad \epsilon_{2T}=(0,0,\textbf{1}_{\rm T}),\nonumber\\
\epsilon_{3L}&=&\frac{1}{\sqrt{2(1-r_2^2)}r_3}(-r_3^2,1-r_2^2,\textbf{0}_{\rm T}), \quad \epsilon_{3T}=(0,0,\textbf{1}_{\rm T}),
\end{eqnarray}
which satisfy the normalization $\epsilon^2_{L}= \epsilon^2_{T}=-1$
and the orthogonality $\epsilon_{2L} \cdot P_2=\epsilon_{3L} \cdot P_3=0$.

For a tensor charmonium, the polarization tensor $\epsilon_{\mu\nu}(\lambda)$ with
helicity $\lambda$ can be constructed via the polarization vector $\epsilon_{\mu}$  \cite{prd82054019,prd83034001}:
\begin{eqnarray}
\epsilon_{\mu\nu}(\pm 2)&=&\epsilon_{\mu}(\pm)\epsilon_{\nu}(\pm), \nonumber\\
\epsilon_{\mu\nu}(\pm 1)&=&\frac{1}{\sqrt{2}}[\epsilon_{\mu}(\pm)\epsilon_{\nu}(0)+\epsilon_{\nu}(\pm)\epsilon_{\mu}(0)], \nonumber\\
\epsilon_{\mu\nu}(0)&=&\frac{1}{\sqrt{6}}[\epsilon_{\mu}(+)\epsilon_{\nu}(-)+\epsilon_{\mu}(-)\epsilon_{\nu}(+)]+
\sqrt{\frac{2}{3}}\epsilon_{\mu}(0)\epsilon_{\nu}(0),
\end{eqnarray}
with $\epsilon(\pm)=\epsilon_{2T}$ and $\epsilon(0)=\epsilon_{2L}$.
It is convenient to define another polarization vector $\epsilon_{\bullet \mu}(\lambda)=
m_2\frac{\epsilon_{\mu\nu(\lambda)}v^{\nu}}{P_2\cdot v}$, which satisfy
\begin{eqnarray}\label{eq:vvv}
\epsilon_{\bullet \mu}(\pm 2)=0,\quad
\epsilon_{\bullet \mu}(\pm 1)= m_2\sqrt{\frac{1}{2}}\frac{\epsilon_{2L}\cdot v}{P_2 \cdot v}\epsilon_{2T\mu},\quad
\epsilon_{\bullet \mu}(0)= m_2\sqrt{\frac{2}{3}}\frac{\epsilon_{2L}\cdot v}{P_2 \cdot v}\epsilon_{2L\mu}.
\end{eqnarray}
The contraction is evaluated as $\frac{\epsilon_{2L}\cdot v}{P_2\cdot v}=\frac{1}{m_2}$ by
neglecting the light meson mass, then we get the relations
$\epsilon_{\bullet T } =\epsilon_{\bullet }(\pm1)= \sqrt{\frac{1}{2}}\epsilon_{2T}$ and
$\epsilon_{\bullet L } =\epsilon_{\bullet }(0)=\sqrt{\frac{2}{3}}\epsilon_{2L}$.
Note that $\epsilon_{\bullet  }$ has the same energy scaling as the
usual polarization vector of a vector meson. It makes the calculations of $B_c$ decays into a tensor meson are similar to
those of vector analogues by replacing the  polarization vector with the corresponding $\epsilon_{\bullet}$.

\subsection{ MESONS WAVE FUNCTION AND THE DISTRIBUTION AMPLITUDES }\label{sec:das}
In the considered decays, there are three typical scales: $M$, $m_2$, and the heavy-meson and heavy-quark mass
difference $\bar{\Lambda}$. They allow for a consistent power expansion in $m_2/M$ and in $\bar{\Lambda}/m_2$ under the hierarchy of
$\bar{\Lambda}\ll m_2\ll M$. In the heavy-quark and large-recoil limits, based on the $k_T$ factorization theorem,
the decay amplitudes are expressed as the convolution of the hard kernels  with the relevant meson wave functions.
The hard kernels can be treated by perturbative QCD at the leading order in an $\alpha_s$ expansion (single gluon exchange
as depicted in Fig. \ref{fig:femy}). The higher-order radiative corrections generate the logarithm divergences, which can be absorbed
into the meson wave functions. One also encounters double logarithm divergences when collinear and soft divergences
overlap, which can be summed to all orders to give a Sudakov factor. After absorbing all the soft dynamics, the initial and
final state meson wave functions can be treated as nonperturbative inputs, which are not calculable but universal.

Analogous  to the $B$ meson \cite{npb591313},
up to first order in $1/M$ under above hierarchy,
the $B_c$ meson wave functions are decomposed into the following Lorentz structures \cite{prd55272,npb5923}:
\begin{eqnarray}
\int d^4z e^{ik_1\cdot z}\langle 0|\bar{b}_{\alpha}(0)c_{\beta}(z)|B_c(P_1)\rangle=
\frac{i}{\sqrt{2N_c}}\{(P_1+M)\gamma_5[\phi_{B_c}(k_1)-\frac{\rlap{/}{n}-\rlap{/}{v}}{\sqrt{2}}\bar{\phi}_{B_c}(k_1)]\}_{\beta\alpha}
\end{eqnarray}
with the two lightlike vectors $n=(1,0,\textbf{0}_{\rm T})$ and $v=(0,1,\textbf{0}_{\rm T})$.
$N_c=3$ is the color factor.
Here, we only consider the  dominant Lorentz structure from the first term, while the second Lorentz structure
 starting from the next-to-leading-power  $\bar{\Lambda}/M$ is numerically neglected \cite{epjc28515,prd76074018}.
 In coordinate space the distribution amplitude $\phi_{B_c}$ is adopted in the form \cite{prd89114019}
 \begin{eqnarray}\label{eq:lcbc}
\phi_{B_c}(x)=N_Bx(1-x)e^{-\frac{m_b+m_c}{8m_bm_c\omega}(\frac{m_c^2}{x}+\frac{m_b^2}{1-x})},
\end{eqnarray}
with the shape parameter $\omega=0.5 \pm 0.1$ GeV   related to  the  factor $N_B$   by the normalization
 \begin{eqnarray}
\int^1_0\phi_{B_c}(x)dx=1.
\end{eqnarray}

For the $P$-wave charmonium states, we use the abbreviations $A$, $S$, and $T$ correspond to axial-vector, scalar,
and tensor charmonium meson, respectively. In terms of
the notation in Ref \cite{prd71114008}, the nonlocal matrix element for the  longitudinally and transversely polarized
axial-vector  and scalar charmonium meson can be decomposed as
\begin{eqnarray}\label{eq:non}
\langle A (P_2,\epsilon_{2L})|\bar{c}_{\alpha}(z)c_{\beta}(0)|0\rangle &=&\frac{1}{\sqrt{2N_c}}\int_0^1dx e^{ixP_2\cdot z}
[m_2\gamma_5\rlap{/}{\epsilon}_{2L} {\psi^L_A(x)}+\gamma_5\rlap{/}{\epsilon}_{2L}\rlap{/}{P}_2 \psi_A^t(x)]_{\beta\alpha},\nonumber\\
\langle A (P_2,\epsilon_{2T})|\bar{c}_{\alpha}(z)c_{\beta}(0)|0\rangle &=&\frac{1}{\sqrt{2N_c}}\int_0^1dx e^{ixP_2\cdot z}
[m_2\gamma_5\rlap{/}{\epsilon}_{2T} { \psi_A^V(x)}+\gamma_5\rlap{/}{\epsilon}_{2T}\rlap{/}{P}_2\psi_A^T(x)]_{\beta\alpha},\nonumber\\
\langle S (P_2)|\bar{c}_{\alpha}(z)c_{\beta}(0)|0\rangle &=&\frac{1}{\sqrt{2N_c}}\int_0^1dx e^{ixP_2\cdot z}
[\rlap{/}{P}_2\psi_S^v(x)+m_2\psi_S^s(x)]_{\beta\alpha},
\end{eqnarray}
where the DAs $\psi^{L,T,v}(x)$ are of twist-2, and $\psi^{t,V,s}(x)$ of twist-3.
As mentioned  in the Introduction,
the charmonium DAs are parameterized using a combination of an universal  factor $\mathcal {T}(x)$ for the $P$-wave states
 and the asymptotic models $ \Phi_{asy}(x)$, given by
\begin{eqnarray}\label{eq:asym}
\psi^{i}\propto \Phi_{asy}^{i}(x)\mathcal {T}(x).
\end{eqnarray}
The expression of $\mathcal {T}(x)$ can be extracted from
$P$-wave Schr$\ddot{o}$dinger states for a Coulomb potential, which are derived
in the Appendix. The asymptotic forms of $\Phi_{asy}^{i}(x)$ in Eq. (\ref{eq:asym})
for the axial-vector mesons can be related to the ones calculated in QCD sum rules by \cite{npb776187}
\begin{eqnarray}\label{eq:phil}
\Phi_A^L(x)=\frac{f_A}{2\sqrt{2N_c}}\phi_{\parallel}(x), \quad \Phi_A^T(x)= \frac{f^{\perp}_A}{2\sqrt{2N_c}}\phi_{\perp}(x),\quad
\Phi_A^t(x)=\frac{f^{\perp}_A}{2\sqrt{2N_c}} h^{(t)}_{\parallel}(x), \quad \Phi_A^V(x)=\frac{f_A}{2\sqrt{2N_c}} g^{(a)}_{\perp}(x),
\end{eqnarray}
where $f_A$ ($f_A^{\perp}$ ) is the vector  (tensor) decay constants.
The leading twist $\phi_{\parallel,\perp}$ can be expanded in a series of Gegenbauer polynomials \cite{npb776187,jhep10108}
\begin{eqnarray}
\phi_{\parallel}&=&6x(1-x)[a_0^{\parallel}+3a_1^{\parallel}(2x-1)+\cdots],\nonumber\\
\phi_{\perp}&=&6x(1-x)[a_0^{\perp}+3a_1^{\perp}(2x-1)+\cdots].
\end{eqnarray}
Owing to the G-parity, $\phi_{\parallel}$ ($\phi_{\perp}$) for ${}^{3}P_1$  state  is symmetric
(antisymmetric) under the exchange of quark and antiquark momentum fractions in the SU(3) limit.
On the contrary, $\phi_{\parallel}$ is antisymmetric for ${}^{1}P_1$ states, while $\phi_{\perp}$
is symmetric in this case.
Thus the  asymptotic forms for twist-2 can be written as
 \begin{eqnarray}\label{eq:twist2}
 \phi_{\parallel}(x)&=&6a_0^{\parallel}x(1-x), \quad  \phi_{\perp}(x)=18a_1^{\perp}x(1-x)(2x-1)   \quad \text{for} \quad {}^{3}P_1; \nonumber\\
 \phi_{\parallel}(x)&=&18a_1^{\parallel}x(1-x)(2x-1), \quad \phi_{\perp}(x)=6a_0^{\perp}x(1-x)  \quad \text{for} \quad {}^{1}P_1.
\end{eqnarray}
Neglecting the three-parton distribution amplitudes containing gluons and terms proportional to
light quark masses,
the twist-3 DAs can be related to the twist-2 ones by Wandzura-Wilczek-type relations \cite{npb776187}:
\begin{eqnarray}
g_{\perp}^{(a)}(x)&=&\frac{1}{2}[\int_0^x du \frac{\phi_{\parallel}(u)}{1-u}+\int_x^1 dv \frac{\phi_{\parallel}(u)}{u}], \nonumber\\
h_{\parallel}^{(t)}(x)&=&(2x-1)[\int_0^x du \frac{\phi_{\perp}(u)}{1-u}-\int_x^1 dv \frac{\phi_{\perp}(u)}{u}],
\end{eqnarray}
which further give
 \begin{eqnarray}\label{eq:twist3}
h_{\parallel}^{(t)}(x)&=&3a_1^{\perp}(2x-1)(1-6x+6x^2), \quad g_{\perp}^{(a)}(x)=\frac{3}{4}a_0^{\parallel}(1+(2x-1)^2)   \quad \text{for} \quad {}^{3}P_1; \nonumber\\
h_{\parallel}^{(t)}(x)&=&3a_0^{\perp}(2x-1)^2, \quad \quad \quad  \quad \quad  g_{\perp}^{(a)}(x)=\frac{3}{2}a_1^{\parallel}(2x-1)^3  \quad \quad \quad \text{for} \quad {}^{1}P_1.
\end{eqnarray}
Combining  Eqs. (\ref{eq:phil}), (\ref{eq:twist2}), and  (\ref{eq:twist3}), we derive
\begin{eqnarray}\label{eq:3p1}
\psi^L(x)&=& \frac{f_A}{2\sqrt{2N_c}}N_Lx(1-x)\mathcal {T}(x),\nonumber\\
\psi^T(x)&=& \frac{f_A^{\perp}}{2\sqrt{2N_c}}N_Tx(1-x)(2x-1)\mathcal {T}(x),\nonumber\\
\psi^t(x)&=& \frac{f_A^{\perp}}{2\sqrt{2N_c}}\frac{N_T}{6}(2x-1)[1-6x+6x^2]\mathcal {T}(x),\nonumber\\
\psi^V(x)&=& \frac{f_A}{2\sqrt{2N_c}}\frac{N_L}{8}[1+(1-2x)^2]\mathcal {T}(x),
\end{eqnarray}
for ${}^{3}P_1$ states, and
\begin{eqnarray}\label{eq:1p1}
\psi^L(x)&=& \frac{f_A}{2\sqrt{2N_c}}N_Tx(1-x)(2x-1)\mathcal {T}(x),\nonumber\\
\psi^T(x)&=& \frac{f_A^{\perp}}{2\sqrt{2N_c}}N_Lx(1-x)\mathcal {T}(x),\nonumber\\
\psi^t(x)&=& \frac{f_A^{\perp}}{2\sqrt{2N_c}}\frac{N_L}{2}(1-2x)^2\mathcal {T}(x),\nonumber\\
\psi^V(x)&=& \frac{f_A}{2\sqrt{2N_c}}\frac{N_T}{12}(2x-1)^3\mathcal {T}(x),
\end{eqnarray}
for  ${}^{1}P_1$ states.
Note that the Gegenbauer moments $a_{0,1}^{\parallel,\perp}$ are absorbed into the coefficients $N_{L,T}$
to satisfy the  normalization conditions \cite{jhep06121}
\begin{eqnarray}
\int_0^1N_Lx(1-x)\mathcal {T}(x)dx=1, \quad
 \int_0^1N_Tx(1-x)(2x-1)^2\mathcal {T}(x)dx=1.
\end{eqnarray}

Because of the charge conjugation invariance, twist-2 and twist-3 DAs of the scalar meson should satisfy
 $\psi_S^v(x)=-\psi_S^v(1-x)$ and $\psi_S^s(x)=\psi_S^s(1-x)$, respectively \cite{pr112173,jhep06067}.
In general, the  asymptotic twist-2 DA can also be expanded in the Gegenbauer polynomials with only odd component contribute \cite{prd67094011,prd74114010}. Based on the description of the charmonium states DAs given above,
$\psi^v_S(x)$ can be recast to the form
\begin{eqnarray}
\psi^v_S(x)&=& \frac{f_S}{2\sqrt{2N_c}}N_Tx(1-x)(2x-1)\mathcal {T}(x).
\end{eqnarray}
As for the  twist-3 DA, we adopt the same  asymptotic form as the pseudoscalar mesons \cite{prd73014017}:
\begin{eqnarray}
\psi^s_S(x)&=& \frac{f_S}{2\sqrt{2N_c}}N_S\mathcal {T}(x),
\end{eqnarray}
with the  normalization condition
\begin{eqnarray}
\int_0^1\psi^s_S(x)dx&=& \frac{f_S}{2\sqrt{2N_c}}.
\end{eqnarray}

The nonlocal matrix element associating with the tensor charmonium can be decomposed as \cite{prd83014008}
\begin{eqnarray}\label{eq:nont}
\langle T (P_2,\epsilon_{\bullet L})|\bar{c}_{\alpha}(z)c_{\beta}(0)|0\rangle &=&\frac{1}{\sqrt{2N_c}}\int_0^1dx e^{ixP_2\cdot z}
[m_2\rlap{/}{\epsilon}_{\bullet L} \psi_T(x)+\rlap{/}{\epsilon}_{\bullet L}\rlap{/}{P}_2 \psi^t_T(x)]_{\beta\alpha},\nonumber\\
\langle T (P_2,\epsilon_{\bullet T})|\bar{c}_{\alpha}(z)c_{\beta}(0)|0\rangle &=&\frac{1}{\sqrt{2N_c}}\int_0^1dx e^{ixP_2\cdot z}
[m_2\rlap{/}{\epsilon}_{\bullet T}  \psi^V_T(x)+\rlap{/}{\epsilon}_{\bullet T}\rlap{/}{P}_2 \psi^T_T(x)]_{\beta\alpha},
\end{eqnarray}
for the longitudinal    and transverse polarizations, respectively.
$\psi_T(x)$, $\psi^T_T(x)$  are leading twist   DAs, and $\psi^V_T(x)$, $\psi^t_T(x)$ are twist-3 ones,
which are related to the ones given in Ref. \cite{prd82054019}
\begin{eqnarray}\label{eq:phi2}
\Phi_T(x)=\frac{f_T}{2\sqrt{2N_c}}\phi_{\parallel}(x), \quad \Phi_T^T(x)= \frac{f^{\perp}_T}{2\sqrt{2N_c}}\phi_{\perp}(x),\quad
\Phi_T^t(x)=\frac{f^{\perp}_T}{2\sqrt{2N_c}} h^{(t)}_{\parallel}(x), \quad \Phi_T^V(x)=\frac{f_T}{2\sqrt{2N_c}} g^{(v)}_{\perp}(x).
\end{eqnarray}
In SU(3) limit, due to the G-parity of the tensor meson,
all of the DAs are antisymmetric under the replacement $x\rightarrow 1-x$.
Here we take the following approximate forms of twist-2 as \cite{prd82054019,prd83014008}
\begin{eqnarray}
\phi_{\parallel}(x)=\phi_{\perp}(x)=N_Tx(1-x)(2x-1),
\end{eqnarray}
and the corresponding expressions for the twist-3 DAs can be derived through the Wandzura-Wilczek relations  as \cite{prd82054019}
\begin{eqnarray}
h^{(t)}_{\parallel}(x)=\frac{N_T}{4}(2x-1)(1-6x+6x^2),\quad g^{(v)}_{\perp}(x)=\frac{N_T}{6}(2x-1)^3.
\end{eqnarray}
Now we can collect the DAs of tensor charmonium states below:
\begin{eqnarray}
\psi_T(x)&=& \frac{f_T}{2\sqrt{2N_c}}N_Tx(1-x)(2x-1)\mathcal {T}(x),\nonumber\\
\psi^T_T(x)&=& \frac{f_T^{\perp}}{2\sqrt{2N_c}}N_Tx(1-x)(2x-1)\mathcal {T}(x),\nonumber\\
\psi^t_T(x)&=& \frac{f_T^{\perp}}{2\sqrt{2N_c}}\frac{N_T}{4}(2x-1)[1-6x+6x^2]\mathcal {T}(x),\nonumber\\
\psi^V_T(x)&=& \frac{f_T}{2\sqrt{2N_c}}\frac{N_T}{6}(2x-1)^3\mathcal {T}(x),
\end{eqnarray}
with the normalization conditions \cite{prd82054019}
\begin{eqnarray}
\int_0^1(2x-1)\psi^{(T)}_T(x)=\frac{f_T^{(\perp)}}{2\sqrt{2N_c}}.
\end{eqnarray}

For the wave functions of light pseudoscalar and vector mesons,
the same forms and parameters are adopted as \cite{prd76074018} and one
is referred to the original literature \cite{ball}.

\subsection{ THE DECAY AMPLITUDES }\label{sec:amp}
In the PQCD approach, the decay amplitudes are expressed
as the convolution of the hard kernels $H$ with the relevant
meson wave functions $\Phi_i$
\begin{eqnarray}
\label{eq:ampu}
\mathcal{A}(B_c\rightarrow M_2M_3) = \int d^4k_1d^4k_2d^4k_3 Tr[C(t)\Phi_1(k_1)\Phi_{2}(k_2)\Phi_3(k_3)
H(k_1,k_2,k_3,t)].
\end{eqnarray}
``Tr'' denotes the trace over all Dirac structures and color indices.
The hadron wave functions $\Phi_i$ absorbed all the nonperturbative components have been described in  Sec. \ref{sec:das}.
The hard kernel $H(k_1,k_2,k_3,t)$ describes the four quark
operator and the spectator quark connected by a hard gluon,
which can be perturbatively calculated including all possible
Feynman diagrams without end-point singularity.
In the following, we start to compute the decay amplitudes of the concerned  decays.
\subsubsection{   Amplitudes for $B_c\rightarrow (S,A,T)P$ decays }
We mark  subscript   $S$, $A$, and $T$ to denote the decay amplitudes contain a
scalar, axial-vector, and tensor charmonium in the final states, respectively.
The amplitudes from factorizable diagrams (a) and (b) in Fig. \ref{fig:femy} for $B_c\rightarrow SP,AP$ decays read as
\begin{eqnarray}\label{eq:fse}
\mathcal{F}_{S}&=&2\sqrt{\frac{2}{3}}C_Ff_Bf_P\pi M^4(r_2^2-1)\int_0^1\int_0^1dx_1dx_2 \int_0^{\infty}\int_0^{\infty}b_1b_2db_1db_2\phi_{B_c}(x_1)
\nonumber\\&&[r_2\psi_S^s(x_2,b_2)(r_b-2x_2)+\psi_S^v(x_2,b_2)(x_2-2r_b)]E_{ab}(t_a)h(\alpha_e,\beta_a,b_1,b_2)S_t(x_2)-
\nonumber\\&&[\psi_S^v(x_2,b_2)(r_c+r_2^2(x_1-1))-2r_2\psi_S^s(x_2,b_2)(r_c+x_1-1)]E_{ab}(t_b)h(\alpha_e,\beta_b,b_2,b_1)S_t(x_1),
\end{eqnarray}
\begin{eqnarray}\label{eq:fae}
\mathcal{F}_{A}&=&-2\sqrt{\frac{2}{3}}C_Ff_Bf_P\pi M^4(r_2^2-1)\int_0^1\int_0^1dx_1dx_2 \int_0^{\infty}\int_0^{\infty}b_1b_2db_1db_2\phi_{B_c}(x_1)
\nonumber\\&&[r_2\psi_A^t(x_2,b_2)(r_b-2x_2)+\psi_A^L(x_2,b_2)(x_2-2r_b)]E_{ab}(t_a)h(\alpha_e,\beta_a,b_1,b_2)S_t(x_2)+
\nonumber\\&&\psi_A^L(x_2,b_2)[r_c+r_2^2(x_1-1)]E_{ab}(t_b)h(\alpha_e,\beta_b,b_2,b_1)S_t(x_1),
\end{eqnarray}
respectively.
The corresponding formula for nonfactorizable   diagrams  (c) and  (d) are
\begin{eqnarray}\label{eq:mse}
\mathcal{M}_{S}&=&\frac{8}{3}C_Ff_B\pi M^4(r_2^2-1)\int_0^1\int_0^1\int_0^1dx_1dx_2dx_3
 \int_0^{\infty}\int_0^{\infty}b_1b_3db_1db_3\phi_{B_c}(x_1)\phi^A_P(x_3)
\nonumber\\&&[\psi_S^v(x_2,b_1)(r_2^2(x_1+2x_2+x_3-2)+x_1-x_3)-r_2\psi_S^s(x_2,b_1)(x_1+x_2-1)]E_{cd}(t_c)h(\beta_c,\alpha_e,b_3,b_1)-
\nonumber\\&&[\psi_S^v(x_2,b_1)(r_2^2(x_2-x_3)+2x_1+x_2+x_3-2)-r_2\psi_S^s(x_2,b_1)(x_1+x_2-1)]E_{cd}(t_d)h(\beta_d,\alpha_e,b_3,b_1),
\end{eqnarray}
\begin{eqnarray}\label{eq:mae}
\mathcal{M}_{A}&=&\frac{8}{3}C_Ff_B\pi M^4(r_2^2-1)\int_0^1\int_0^1\int_0^1dx_1dx_2dx_3
 \int_0^{\infty}\int_0^{\infty}b_1b_3db_1db_3\phi_{B_c}(x_1)\phi^A_P(x_3)
\nonumber\\&&[\psi_A^L(x_2,b_1)(r_2^2-1)(x_1-x_3)-r_2\psi_A^t(x_2,b_1)(x_1+x_2-1)]E_{cd}(t_c)h(\beta_c,\alpha_e,b_3,b_1)+
\nonumber\\&&[\psi_A^L(x_2,b_1)(r_2^2(x_2-x_3)+2x_1+x_2+x_3-2)-r_2\psi_A^t(x_2,b_1)(x_1+x_2-1)]E_{cd}(t_d)h(\beta_d,\alpha_e,b_3,b_1),
\end{eqnarray}
with $r_{b,c}=m_{b,c}/M$. $C_F=4/3$ is a color factor. $f_P$ is the decay constant of the light pseudoscalar meson, emitted from the weak vertex.
 The functions $h$, $E$ and the factorization scales $t_{a,b,c,d}$ can be found in \cite{rui1}.
 The leading twist DAs of the pseudoscalar meson $\phi_P^A$ and
 the jet function $S_t(x)$ come from  \cite{prd76074018}.
$\alpha_e$ and $\beta_{a,b,c,d}$ are the virtuality of the internal gluon and quarks, respectively. Their expressions are
\begin{eqnarray}
\alpha_e&=&-[x_1-(1-x_2)r_2^2](x_1+x_2-1)M^2,\nonumber\\
\beta_a&=&[r^2_b-x_2(1+r_2^2(x_2-1))]M^2,\nonumber\\
\beta_b&=&[r_c^2+(x_1-1)(r_2^2-x_1)]M^2,\nonumber\\
\beta_c&=&-(x_1+x_2-1)[x_1-x_3+r_2^2(x_2+x_3-1)]M^2,\nonumber\\
\beta_d&=&-(x_1+x_2-1)[x_1+x_3-1+r_2^2(x_2-x_3)]M^2.
\end{eqnarray}
It should be stressed that    the nonlocal matrix element for the axial-vector and  scalar  charmonium meson in Eq. (\ref{eq:non})
can be  related to the vector and pseudoscalar ones \cite{rui1,rui2}, respectively,
 by multiplying by the structure $-(i)\gamma_5$ from the left hand.
The factorization formulas ($\mathcal{F/M}$) here and below are similar to the corresponding ones in \cite{rui1,rui2} with some terms flipping signs.
As mentioned before, the nonlocal matrix element associating with the tensor charmonium in Eq. (\ref{eq:nont})
is also analogous to the vector case,  except that the polarization vector is replaced by $\epsilon_{\bullet  }$.
Therefore  the correspondence between a tensor meson and a axial-vector
meson allows us to get the factorization formulas of $B_c\rightarrow TP$ as
\begin{eqnarray}\label{eq:fte}
\mathcal{F}_{T}=\sqrt{\frac{2}{3}}\mathcal{F}_{A}|_{\psi_A^L\rightarrow \psi_T, \psi_A^t\rightarrow \psi_T^t,r_c\rightarrow -r_c },
\quad
\mathcal{M}_{T}=\sqrt{\frac{2}{3}}\mathcal{M}_{A}|_{\psi_A^L\rightarrow \psi_T, \psi_A^t\rightarrow \psi_T^t}.
\end{eqnarray}

With the functions obtained in the above, the total decay amplitudes  for the $B_c\rightarrow (S,A,T)P$ are given by
\begin{eqnarray}\label{eq:satp}
\mathcal{A}(B_c\rightarrow (S,A,T)P)=V^{*}_{cb}V_{uq}[(C_2+\frac{1}{3}C_1)\mathcal{F}_{S,A,T}+C_1\mathcal{M}_{S,A,T}].
\end{eqnarray}
\subsubsection{  Amplitudes for $B_c\rightarrow (S,A,T)V$ decays }
For $B_c\rightarrow SV$ decays, the  decay amplitudes of factorization emission diagrams and nonfactorization emission diagrams are given as
\begin{eqnarray}\label{eq:fsev}
\mathcal{F}_{S}&=&2\sqrt{\frac{2}{3}}C_Ff_Bf_V\pi M^4\sqrt{1-r_2^2}\int_0^1\int_0^1dx_1dx_2 \int_0^{\infty}\int_0^{\infty}b_1b_2db_1db_2\phi_{B_c}(x_1)
\nonumber\\&&[r_2\psi_S^s(x_2,b_2)(r_b-2x_2)+\psi_S^v(x_2,b_2)(x_2-2r_b)]E_{ab}(t_a)h(\alpha_e,\beta_a,b_1,b_2)S_t(x_2)-
\nonumber\\&&[\psi_S^v(x_2,b_2)(r_c+r_2^2(x_1-1))-2r_2\psi_S^s(x_2,b_2)(r_c+x_1-1)]E_{ab}(t_b)h(\alpha_e,\beta_b,b_2,b_1)S_t(x_1),
\end{eqnarray}
\begin{eqnarray}\label{eq:msev}
\mathcal{M}_S&=&\frac{8}{3}C_Ff_B\pi M^4\sqrt{1-r_2^2}\int_0^1\int_0^1\int_0^1dx_1dx_2dx_3
 \int_0^{\infty}\int_0^{\infty}b_1b_3db_1db_3\phi_{B_c}(x_1)\phi_V(x_3)
\nonumber\\&&[\psi_S^v(x_2,b_1)(r_2^2(x_1+2x_2+x_3-2)+x_1-x_3)-r_2\psi_S^s(x_2,b_1)(x_1+x_2-1)]E_{cd}(t_c)h(\beta_c,\alpha_e,b_3,b_1)-
\nonumber\\&&[\psi_S^v(x_2,b_1)(r_2^2(x_2-x_3)+2x_1+x_2+x_3-2)-r_2\psi_S^s(x_2,b_1)(x_1+x_2-1)]E_{cd}(t_d)h(\beta_d,\alpha_e,b_3,b_1),
\end{eqnarray}
where $f_V$ and $\phi_V$ are the decay constants and the twist-2 distribution amplitudes of the light vector mesons, respectively.
The total decay amplitudes for $B_c\rightarrow SV$ decays are  similar to that of $B_c\rightarrow SP$ in  Eq. (\ref{eq:satp}) with the replacement
$f_P\rightarrow f_V$, $\phi_P^A\rightarrow \phi_V$.

Like vector mesons, axial-vector mesons  also carry spin degrees of freedom. Therefore,
$B_c\rightarrow AV$ decays contain more amplitudes associated with three
different polarizations. 
We mark  superscript $L$, $N$ and $T$ to denote the contributions from longitudinal polarization, normal polarization, and transverse
polarization, respectively.
\begin{eqnarray}\label{eq:faevl}
\mathcal{F}^{L}_A&=&2\sqrt{\frac{2}{3}}C_Ff_Bf_V\pi M^4\sqrt{1-r_2^2}\int_0^1\int_0^1dx_1dx_2 \int_0^{\infty}\int_0^{\infty}b_1b_2db_1db_2\phi_{B_c}(x_1)
\nonumber\\&&[r_2\psi_A^t(x_2,b_2)(r_b-2x_2)+\psi_A^L(x_2,b_2)(x_2-2r_b)]E_{ab}(t_a)h(\alpha_e,\beta_a,b_1,b_2)S_t(x_2)+
\nonumber\\&&\psi_A^L(x_2,b_2)[r_c+r_2^2(x_1-1)]E_{ab}(t_b)h(\alpha_e,\beta_b,b_2,b_1)S_t(x_1),
\end{eqnarray}
\begin{eqnarray}\label{eq:faevn}
\mathcal{F}^{N}_A&=&2\sqrt{\frac{2}{3}}C_Ff_Bf_Vr_3\pi M^4\int_0^1\int_0^1dx_1dx_2 \int_0^{\infty}\int_0^{\infty}b_1b_2db_1db_2\phi_{B_c}(x_1)
\nonumber\\&&[\psi_A^T(x_2,b_2)(r_2^2(r_b+2-4x_2)+r_b-2)+r_2\psi_A^V(x_2,b_2)
(r_2^2(x_2-1)+x_2+1)-4r_b]\nonumber\\&&E_{ab}(t_a)h(\alpha_e,\beta_a,b_1,b_2)S_t(x_2)-
\psi_A^V(x_2,b_2)r_2[1-2r_c-2x_1+r_2^2]E_{ab}(t_b)h(\alpha_e,\beta_b,b_2,b_1)S_t(x_1),
\end{eqnarray}
\begin{eqnarray}\label{eq:faevt}
\mathcal{F}^{T}_A&=&2\sqrt{\frac{2}{3}}C_Ff_Bf_Vr_3\pi M^4(r_2^2-1)\int_0^1\int_0^1dx_1dx_2 \int_0^{\infty}\int_0^{\infty}b_1b_2db_1db_2\phi_{B_c}(x_1)
\nonumber\\&&[\psi_A^T(x_2,b_2)(2-r_b)+r_2\psi_A^V(x_2,b_2)
(x_2-1)]\nonumber\\&&E_{ab}(t_a)h(\alpha_e,\beta_a,b_1,b_2)S_t(x_2)+
\psi_A^V(x_2,b_2)r_2E_{ab}(t_b)h(\alpha_e,\beta_b,b_2,b_1)S_t(x_1),
\end{eqnarray}
\begin{eqnarray}\label{eq:maevl}
\mathcal{M}^{L}_A&=&\frac{8}{3}C_Ff_B\pi M^4\sqrt{1-r_2^2}\int_0^1\int_0^1\int_0^1dx_1dx_2dx_3
 \int_0^{\infty}\int_0^{\infty}b_1b_3db_1db_3\phi_{B_c}(x_1)\phi_V(x_3)
\nonumber\\&&[-\psi_A^L(x_2,b_1)(r_2^2-1)(x_1-x_3)+r_2\psi_A^t(x_2,b_1)(x_1+x_2-1)]E_{cd}(t_c)h(\beta_c,\alpha_e,b_3,b_1)-
\nonumber\\&&[\psi_A^L(x_2,b_1)(r_2^2(x_2-x_3)+2x_1+x_2+x_3-2)-r_2\psi_A^t(x_2,b_1)(x_1+x_2-1)]E_{cd}(t_d)h(\beta_d,\alpha_e,b_3,b_1),
\end{eqnarray}
\begin{eqnarray}\label{eq:maevn}
\mathcal{M}^{N}_A&=&\frac{8}{3}C_Ff_Br_3\pi M^4\int_0^1\int_0^1\int_0^1dx_1dx_2dx_3
 \int_0^{\infty}\int_0^{\infty}b_1b_3db_1db_3\phi_{B_c}(x_1)
\nonumber\\&&[\phi_V^a(x_3)\psi_A^V(x_2,b_1)(r_2^2-1)(x_2+x_3-1)
+\phi_V^v(x_3)\psi_A^T(x_2,b_1)(r_2^2(x_1+2x_2+x_3-2)+x_1-x_3)]\nonumber\\&&E_{cd}(t_c)h(\beta_c,\alpha_e,b_3,b_1)+
[\phi_V^a(x_3)2(r_2^2-1)(r_2(x_2-x_3)\psi_A^V(x_2,b_1)+2(x_1+x_3-1)
\psi_A^T(x_2,b_1))\nonumber\\&&+\phi_V^v(x_3)(2r_2\psi_A^V(x_2,b_1)(r_2^2(x_2-x_3)+2x_1+x_2+x_3-2)
+\nonumber\\&&\psi_A^T(x_2,b_1)(r_2^2(1-x_1-2x_2+x_3)+1-x_1-x_3))]E_{cd}(t_d)h(\beta_d,\alpha_e,b_3,b_1),
\end{eqnarray}
\begin{eqnarray}\label{eq:maevt}
\mathcal{M}^{T}_A&=&-\frac{8}{3}C_Ff_Br_3\pi M^4\int_0^1\int_0^1\int_0^1dx_1dx_2dx_3
 \int_0^{\infty}\int_0^{\infty}b_1b_3db_1db_3\phi_{B_c}(x_1)
\nonumber\\&&[\phi_V^a(x_3)\psi_A^V(x_2,b_1)2r_2(r^2_2(x_2+x_3-1)+2x_1+x_2-x_3-1)
+\phi_V^v(x_3)\psi_A^T(x_2,b_1)(r_2^2-1)(x_1-x_3)]\nonumber\\&&E_{cd}(t_c)h(\beta_c,\alpha_e,b_3,b_1)+
[2\phi_V^a(x_3)(r_2\psi_A^V(x_2,b_1)(r_2^2(x_2-x_3)+2x_1+x_2+x_3-2)\nonumber\\&&+2
\psi_A^T(x_2,b_1)(r_2^2(1-x_1-2x_2+x_3)+1-x_1-x_3))\nonumber\\&&+\phi_V^v(x_3)\psi_A^T(x_2,b_1)
(r_2^2-1)(x_1+x_3-1)]E_{cd}(t_d)h(\beta_d,\alpha_e,b_3,b_1),
\end{eqnarray}
where $\phi_V^a$ and $\phi_V^v$ are the two twist-3 distribution amplitudes for the transverse  polarization of light vector mesons.
For $B_c\rightarrow TV$ decays, the  decay amplitudes can be related to the axial-vector ones by making the following replacement:
\begin{eqnarray}
\mathcal{F(M)}^{L}_T&=&\sqrt{\frac{2}{3}}\mathcal{F(M)}^{L}_A|_{\psi_A^L\rightarrow \psi_T, \psi_A^t\rightarrow \psi_T^t,r_c\rightarrow -r_c },\nonumber\\ \mathcal{F(M)}^{N,T}_T&=&\sqrt{\frac{1}{2}}\mathcal{F(M)}^{N,T}_A|_{\psi_A^T\rightarrow \psi_T^T, \psi_A^V\rightarrow \psi_T^V,r_c\rightarrow -r_c },
\end{eqnarray}
 where the factors $\sqrt{\frac{2}{3}}$ and $\sqrt{\frac{1}{2}}$ come from the equivalent polarization vector
  $\epsilon_{\bullet }$ in Eq. (\ref{eq:vvv}) of the tensor mesons for  the longitudinal and transverse  polarizations, respectively.
  The total decay amplitudes for $B_c\rightarrow (A,T)V$ decays can be decomposed as
\begin{eqnarray}\label{eq:atv1}
\mathcal{A}(B_c\rightarrow (A,T)V)=\mathcal{A}^L_{A,T}+\mathcal{A}^N_{A,T}\epsilon_{2T}\cdot\epsilon_{3T}
+i\mathcal{A}^T_{A,T}\epsilon_{\alpha\beta\rho\sigma}n^{\alpha}v^{\beta}\epsilon_{2 T}^{\rho}\epsilon_{3 T}^{\sigma},
\end{eqnarray}
where the three  polarization amplitudes   have the same structure as Eq. (\ref{eq:satp}).

\section{Numerical results}\label{sec:results}
To proceed the numerical analysis, it is useful to summarize all of the input quantities we have used in this work.
The central values (in GeV) of the relevant meson masses and heavy quark masses are adopted as  \cite{pdg2016}
\begin{eqnarray}
M&=&6.275, \quad m_b=4.8, \quad m_c=1.275, \quad m_{\rho}=0.775, \quad m_{K^*}=0.892.
\end{eqnarray}
While the masses of the $P$-wave charmonium have been given  in Table \ref{tab:properties}.
The CKM matrix-elements are set as $|V_{cb}|=0.0405$, $|V_{us}|=0.2248$, and
$|V_{ud}|=0.97417$  \cite{pdg2016}.
For the decay constants of $P$-wave charmonium, the detailed discussions
in the nonrelativistic QCD factorization, the light-front approach and the QCD sum rules,
 could be found in Refs. \cite{prd79074004,prd96014026,jhep10074,jhep06121}.
Here we employ the most recent updated values (in GeV) evaluated from the QCD sum rules at the scale $\mu=m_c$ \cite{prd96014026}:
\begin{eqnarray}\label{eq:fpwave}
f_{\chi_{c0}}&=&0.0916, \quad f_{\chi_{c1}}=0.185, \quad f_{\chi_{c2}}=0.177,\quad  f_{h_{c}}=0.127, \nonumber\\
\quad f^{\perp}_{\chi_{c1}}&=&0.0875, \quad f^{\perp}_{\chi_{c2}}=0.128, \quad  f^{\perp}_{h_{c}}=0.133.
\end{eqnarray}
For the decay constants of light mesons, we use \cite{prd76074018}
\begin{eqnarray}
f_{\pi}&=&0.131 ,\quad f_{K}=0.160 ,\quad f_{\rho}=0.209  ,\nonumber\\ f_{K^*}&=&0.217  ,\quad
\quad f_{\rho}^{\perp}=0.165  ,\quad f_{K^*}^{\perp}=0.185 \quad \text{GeV}.
\end{eqnarray}
The $B_c$ meson decay constant and lifetime  are adopted as  $f_{B_{c}}=0.489$ GeV \cite{rui1,rui2}
and $\tau_{B_c}=0.507$ ps \cite{pdg2016}, respectively.

The branching ratios for the considered  decays in the $B_c$ meson rest frame can be written as
\begin{eqnarray}
\mathcal {B}=\frac{G^2_F\tau_{B_c}}{32\pi M}(1-r^2_2)|\mathcal {A}|^2,
\end{eqnarray}
where  the decay amplitudes $\mathcal {A}$ for
each channel have been given explicitly in the previous section.
When the final states involve   axial-vector/tensor charmonium states and a vector meson, the decay
amplitude can be decomposed into three components,
\begin{eqnarray}
|\mathcal {A}|^2=|\mathcal {A}_0|^2+|\mathcal {A}_{\parallel}|^2+|\mathcal {A}_{\perp}|^2,
\end{eqnarray}
where $\mathcal {A}_0, \mathcal {A}_{\parallel},\mathcal {A}_{\perp}$ refer to the longitudinal,
parallel, and perpendicular polarization amplitudes in the transversity
basis, respectively, which are related to $\mathcal {A}^{L,N,T}$ of Eq.~(\ref{eq:atv1}) via
\begin{eqnarray}\label{eq:spin}
\mathcal {A}_0=\mathcal {A}^L, \quad \mathcal {A}_{\parallel}=\sqrt{2}\mathcal {A}^{N}, \quad \mathcal {A}_{\perp}=\sqrt{2}\mathcal {A}^{T}.
\end{eqnarray}

\begin{table}
\caption{The PQCD predictions on 
branching ratios of $B_c$ decays to final states containing a $P$-wave charmonium
state and a light pseudoscalar meson.  The errors for these entries
correspond to the uncertainties in hadronic shape parameters, from the decay constants, and the scale dependence, respectively.
For  comparison, we also list  other theoretical results.
Note that some branching ratios are evaluated with the Wilson coefficient $a_1=1.14$ in the referred models.}
\label{tab:br1}
\begin{tabular}[t]{l cccccccc}
\hline\hline
Modes     & This work &\cite{prd82034019} &\cite{prd74074008}&\cite{prd73054024}
 &\cite{prd65014017} &\cite{jpg28595} &\cite{jpg39015009} &\cite{171007011} \\ \hline
$B^+_c \rightarrow \chi_{c0} \pi^+ $ &$(1.6^{+0.2+0.3+0.0}_{-0.2-0.3-0.1})\times 10^{-3}$ &$2.1\times 10^{-4}$ &$2.6\times 10^{-4}$
  &$5.5\times 10^{-4}$  &$2.8\times 10^{-4}$ &$9.8\times 10^{-3}$  &$3.1\times 10^{-4}$  &$4.2\times 10^{-3}$   \\
$B^+_c \rightarrow \chi_{c1} \pi^+ $ &$(5.1^{+0.3+1.1+0.0}_{-0.4-1.1-0.2})\times 10^{-4}$ &$2.0\times 10^{-4}$ &$1.4\times 10^{-6}$
 &$6.8\times 10^{-5}$ &$7.0\times 10^{-5}$ &$8.9\times 10^{-5}$   &$2.1\times 10^{-5}$   &$5.0\times 10^{-5}$    \\
$B^+_c \rightarrow \chi_{c2} \pi^+ $ &$(4.0^{+0.3+0.9+0.3}_{-0.4-0.8-0.3})\times 10^{-3}$ &$3.8\times 10^{-4}$ &$2.2\times 10^{-4}$
 &$4.6\times 10^{-4}$ &$2.5\times 10^{-4}$ &$7.9\times 10^{-3}$ &$2.1\times 10^{-4}$     &$7.4\times 10^{-4}$    \\
$B^+_c \rightarrow h_c \pi^+ $       &$(5.4^{+0.4+1.0+0.4}_{-0.3-1.0-0.3})\times 10^{-4}$ &$4.6\times 10^{-4}$ &$5.3\times 10^{-4}$
 &$1.1\times 10^{-3}$    &$5.0\times 10^{-4}$ &$1.6\times 10^{-2}$  &$9.8\times 10^{-4}$  &$6.2\times 10^{-3}$    \\
 $B^+_c \rightarrow \chi_{c0} K^+ $       &$(1.2^{+0.2+0.3+0.0}_{-0.2-0.2-0.1})\times 10^{-4}$ &$1.6\times 10^{-5}$ &$2.0\times 10^{-5}$
 &$4.2\times 10^{-5}$    &$2.1\times 10^{-6}$ &--  &$2.3\times 10^{-5}$  &$3.2\times 10^{-4}$    \\
 $B^+_c \rightarrow \chi_{c1} K^+ $      &$(3.8^{+0.3+0.9+0.1}_{-0.3-0.8-0.1})\times 10^{-5}$ &$1.5\times 10^{-5}$ &$1.1\times 10^{-7}$
 &$5.1\times 10^{-6}$ &$5.2\times 10^{-7}$ &--  &$1.6\times 10^{-6}$   &$4.0\times 10^{-6}$    \\
$B^+_c \rightarrow \chi_{c2} K^+ $       &$(3.1^{+0.3+0.7+0.2}_{-0.2-0.6-0.2})\times 10^{-4}$ &$2.8\times 10^{-5}$ &$1.7\times 10^{-5}$
 &$3.4\times 10^{-5}$ &$1.8\times 10^{-6}$ &-- &$1.6\times 10^{-5}$     &$5.6\times 10^{-5}$    \\
$B^+_c \rightarrow h_c K^+ $             &$(4.3^{+0.3+0.7+0.3}_{-0.2-0.8-0.2})\times 10^{-5}$ &$3.5\times 10^{-5}$ &$4.1\times 10^{-5}$
&$8.3\times 10^{-5}$ &$3.8\times 10^{-6}$ &--  &$7.4\times 10^{-5}$  &$4.7\times 10^{-4}$    \\
\hline\hline
\end{tabular}
\end{table}

\begin{table}
\caption{The PQCD predictions on
branching ratios  of $B_c$ decays to final states containing a $P$-wave charmonium
state and a light vector meson.  The errors for these entries
correspond to the uncertainties in hadronic shape parameters, from the decay constants, and the scale dependence, respectively.
For  comparison, we also list  other theoretical results.
Note that some branching ratios are evaluated with the Wilson coefficient $a_1=1.14$ in the referred models.}
\label{tab:br2}
\begin{tabular}[t]{l cccccccc}
\hline\hline
Modes     & This work &\cite{prd82034019} &\cite{prd74074008}&\cite{prd73054024}
 &\cite{prd65014017} &\cite{jpg28595} &\cite{jpg39015009} &\cite{prd87034004} \\ \hline
$B^+_c \rightarrow \chi_{c0} \rho^+ $&$(5.8^{+0.6+1.1+0.4}_{-0.6-1.2-0.4})\times 10^{-3}$   &$5.8\times 10^{-4}$ &$6.7\times 10^{-4}$
 &$1.3\times 10^{-3}$    &$7.2\times 10^{-4}$  &$3.3\times 10^{-2}$ &$7.6\times 10^{-4}$  &--   \\
$B^+_c \rightarrow \chi_{c1} \rho^+ $&$(2.8^{+0.2+0.5+0.1}_{-0.2-0.5-0.1})\times 10^{-3}$   &$1.5\times 10^{-4}$ &$1.0\times 10^{-4}$
 &$2.9\times 10^{-4}$    &$2.9\times 10^{-4}$ &$4.6\times 10^{-3}$ &$2.3\times 10^{-4}$  & $1.47\times 10^{-3}$ \\
$B^+_c \rightarrow \chi_{c2} \rho^+ $&$(1.6_{-0.1-0.3-0.0}^{+0.1+0.4+0.1})\times 10^{-2}$   &$1.1\times 10^{-3}$  &$6.5\times 10^{-4}$
 &$1.2\times 10^{-3}$ &$5.1\times 10^{-4}$ &$3.2\times 10^{-2}$ &$5.6\times 10^{-4}$&--\\
$B^+_c \rightarrow h_c  \rho^+ $    &$(2.3^{+0.1+0.3+0.2}_{-0.2-0.4-0.1})\times 10^{-3}$   &$1.0\times 10^{-3}$  &$1.3\times 10^{-3}$
 &$2.5\times 10^{-3}$ &$1.2\times 10^{-3}$ &$5.3\times 10^{-2}$ &$2.2\times 10^{-3}$  &$1.24\times 10^{-3}$          \\
 $B^+_c \rightarrow \chi_{c0} K^{*+} $&$(3.3^{+0.4+0.7+0.2}_{-0.3-0.6-0.2})\times 10^{-4}$   &$4.0\times 10^{-5}$ &$3.7\times 10^{-5}$
 &$7.0\times 10^{-5}$    &$3.9\times 10^{-6}$  &-- &$4.5\times 10^{-5}$  &--   \\
$B^+_c \rightarrow \chi_{c1} K^{*+} $&$(1.8^{+0.2+0.3+0.3}_{-0.1-0.3-0.1})\times 10^{-4}$   &$1.0\times 10^{-5}$ &$7.3\times 10^{-6}$
 &$1.8\times 10^{-5}$    &$1.8\times 10^{-6}$ &-- &$1.7\times 10^{-5}$  & $7.07\times 10^{-5}$ \\
$B^+_c \rightarrow \chi_{c2} K^{*+} $&$(9.6^{+0.7+2.0+0.6}_{-0.8-1.8-0.4})\times 10^{-4}$   &$7.4\times 10^{-5}$  &$3.8\times 10^{-5}$
 &$6.5\times 10^{-5}$ &$3.1\times 10^{-6}$ &--&$3.3\times 10^{-5}$&--\\
$B^+_c \rightarrow h_c  K^{*+} $    &$(1.3^{+0.1+0.3+0.2}_{-0.1-0.2-0.0})\times 10^{-4}$   &$7.0\times 10^{-5}$  &$7.1\times 10^{-5}$
 &$1.3\times 10^{-4}$ &$6.8\times 10^{-6}$ &-- &$1.3\times 10^{-4}$  &$6.18\times 10^{-5}$          \\
\hline\hline
\end{tabular}
\end{table}

Our numerical results of branching ratios for $B_c\rightarrow (S,A,T)P$  and $B_c\rightarrow (S,A,T)V$
decays are listed in Tables \ref{tab:br1} and \ref{tab:br2}, respectively.
The first kind of uncertainties is from the shape parameter
$\omega$ in the  wave function of the $B_c$ meson and the charm-quark mass $m_c$.
 In the evaluation, we vary the value of $\omega$ within a $20\%$ range and $m_c = 1.275$ GeV by $ \pm 0.025$ GeV.
The second error comes from the decay constants of the $P$-wave charmonium meson in Eq. (\ref{eq:fpwave}), which
varies $10\%$ for error estimates.
The last one is caused  by the hard scale $t$ located between $0.75\sim 1.25$
times the invariant masses of the internal particles, which characterizes the  size of higher-order corrections to the hard amplitudes.
 It turns out that the errors are dominant by the uncertainties from the decay
 constants of $P$-wave charmonium meson distribution amplitudes, which can reach $20\%$ in magnitude.
As discussed in Ref \cite{epjc76564}, by using the light-cone wave function for
the $B_c$  meson, the theoretical uncertainty from the 
the charm-quark mass is largely reduced. It is also found that the branching ratios  are insensitive to
the hard scale, which  means the higher-order contributions can be safely neglected.
In a recent paper \cite{jhep12012}, the authors claimed that the relativistic corrections to light-cone distribution
amplitudes of $S$-wave heavy quarkonia  are comparable with the next-to-leading order radiative corrections.
In view of this point,
we  check  the sensitivity of our results to the   squared velocity
 $v^2$ of the charm quark inside the $P$-wave charmonium  states in Eq.(\ref{eq:vv2}).
The variation of $v^2$ in the range $0.25\sim 0.35$ will result in the branching ratios changing only  a few percents.
This is similar to the comment in \cite{prd77054003} that
 the relativistic  corrections to the Coulomb wave functions would be less significant.
In addition, the uncertainties related to the light  mesons,
such as the  decay constants and the Gegenbauer
moments shown in \cite{prd76074018}, are less than $10\%$.
Therefore they have been neglected in our calculations.

It can be seen that
the  former four processes ( including one $\pi$ or $\rho$ meson in the  final states)
 have a relatively large branching ratios   owing to the CKM factor enhancement, while
the branching ratios of the latter four processes ( including one $K$ or $K^*$ meson in the final states)
are comparatively small  due to the CKM factor suppression.
Since the two type decays  have identical topology and similar kinematic properties.
In the limit of SU(3) flavor symmetry, the relative
ratios  $\mathcal{R}_{K/\pi}\equiv\mathcal{B}(B_c\rightarrow (S,A,T)K)/\mathcal{B}(B_c\rightarrow (S,A,T)\pi)$
and  $\mathcal{R}_{K^{*}/\rho}\equiv\mathcal{B}(B_c\rightarrow (S,A,T)K^*)/\mathcal{B}(B_c\rightarrow (S,A,T)\rho)$
are dominated by the ratio of the relevant CKM matrix
elements $|V_{us}|^2/|V_{ud}|^2\sim \lambda^2$ under the naive
factorization approximation. After including  the kaon ($K^*$) and pion ($\rho$) decay constants,
one expects $\mathcal{R}_{K/\pi}\sim 0.081$  and $\mathcal{R}_{K^{*}/\rho}\sim 0.057$.
From Tables \ref{tab:br1} and \ref{tab:br2},
our  predictions for $\mathcal{R}_{K/\pi}$ corresponding to various $P$-wave charmonium states  lie in the range   0.075 to 0.080, while $\mathcal{R}_{K^{*}/\rho}$  
is in the range 0.057 to 0.064,  both are very close to the above expected values.
It means that the dominant contributions to the branching ratios come from the factorizable
topology, while the nonfactorizable contribution is suppressed by the Wilson coefficient $C_1$
[see Eq. (\ref{eq:satp})]. 

One can see some  interesting hierarchical relations among these branching ratios in our predictions. For example,
branching ratios for decays involving  pseudoscalar mesons
in the final state are smaller than their vector partners for the same flavor content.
This is partially due to the pseudoscalar meson decay constant is usual smaller than the vector ones.
Furthermore, since the $B_c$ meson is a spinless particle,
 according to the  angular momentum conservation,
only one partial wave contribute to the $B_c\rightarrow SP, AP, TP, SV$ decays,
while in the $AV, TV$ modes,  three partial waves are simultaneously allowed, resulting in the larger branching ratios. 
For those channels with the  same light meson and different $P$-wave charmonium mesons
 in the  final states, we have the following hierarchy pattern:
\begin{eqnarray}\label{eq:hie}
\mathcal {B}(B_c\rightarrow \chi_{c2}P)>\mathcal {B}(B_c\rightarrow \chi_{c0}P)>
\mathcal {B}(B_c\rightarrow \chi_{c1}P)\sim \mathcal {B}(B_c\rightarrow h_c P).
\end{eqnarray}
As discussed   in Ref \cite{rui2}, the branching ratio of $B_c\rightarrow \eta_c(2S) \pi$ is enhanced by the
the twist-3 distribution amplitude from Fig. 3(b). Nevertheless, 
this contribution vanishes   for  $B_c\rightarrow \psi(2S)\pi$ decay
since the Lorentz structure of the vector charmonium wave functions is different from the pseudoscalar case.
As mentioned in the previous section,
because of the corresponding relation between a pseudoscalar (vector) and a scalar (axial-vector) charmonium,
the similar situation also exists in this work.
The twist-3 distribution amplitude  from Fig \ref{fig:femy}(b) also give the dominant contribution to the
$B_c\rightarrow \chi_{c0}P$ decays, while for other channels, the dominant contribution still come from the twist-2 ones.
Because the strong interference between the twist-2 and twist-3 contributions   is constructive in $B_c\rightarrow \chi_{c0}P$,
we   have a large branching ratio for this mode.
One can see that the dominant twist-2 contributions for $B_c\rightarrow (\chi_{c1},h_c) P$
 are suppressed by a factor of $r_c-r^2_2$ given in Eq. (\ref{eq:fae}),
whereas this  suppression  is absent in the case of $B_c\rightarrow \chi_{c2} P$
 due to the $r_c$ term flipping sign [see Eq. (\ref{eq:fte})].
 This explains why  $B_c\rightarrow \chi_{c2}P$ has a rate greater than $\chi_{c1}P$ and $h_cP$.
Of course,  this is only a rough estimate on the magnitudes, the branching ratios also have been related
to the decay constants and the distribution amplitudes of the various $P$-wave charmonium mesons.
The  relations between the decay  constants $f_{\chi_{c2}}>f_{\chi_{c0}}$ in Eq. (\ref{eq:fpwave}) implies that
$\mathcal {B}(B_c\rightarrow \chi_{c2}P)>\mathcal {B}(B_c\rightarrow \chi_{c0}P)$.
The similar pattern also occurs $B_c\rightarrow (A,S,T)V$ decays; see Table \ref{tab:br2}.

As mentioned in the Introduction,  many   other work have performed  a systematic study on
the $P$-wave charmonium decays of $B_c$ mesons. Various approaches such as
several relativistic and nonrelativistic quark models \cite{prd82034019,prd74074008,prd73054024,prd65014017},
the  sum rules of QCD \cite{jpg28595},
the improved Bethe-Salpeter approach \cite{jpg39015009}, the Isgur-Scora-Grinstein-Wise II model \cite{prd87034004},
 and the nonrelativistic QCD effective theory \cite{171007011}
have been used to calculate the branching ratios.
For the sake of comparison, we  briefly list the obtained  theoretical results in Tables \ref{tab:br1} and \ref{tab:br2}.
One finds that some of  the results  given by   different models are roughly comparable.
For example, our theoretical predictions on those decays involving the $h_c$ meson in the final state
are of the same order of magnitude as observed in \cite{prd82034019,prd74074008,prd73054024,jpg39015009}.
The branching ratios of $B_c\rightarrow \chi_{c1}\pi$ and
 $B_c\rightarrow h_{c}\pi$ evaluated by N. Sharma {\it et al.} \cite{jpg37075013}
 are $7\times 10^{-4}$ and $6\times 10^{-4}$, respectively,  which also match well with our results.
 In a very recent paper \cite{170900550}, the author predicted the branching ratio
$\mathcal {B}(B_c^+\rightarrow \chi_{c0}\pi^+)=1.22\times 10^{-3}$, which is comparable
to our prediction. Of course, some predicted values are quite a spread in various models.
The predictions in Ref \cite{jpg28595} are typically larger excepted for $B_c^+\rightarrow \chi_{c1}\pi^+$.
Previously, G. L. Castro {\it et al.} \cite{jpg282241} studied the nonleptonic decays of the $B_c$ into tensor mesons using
the factorization hypothesis. They predict $\mathcal {B}(B_c\rightarrow \chi_{c2}\pi)$,  $\mathcal {B}(B_c\rightarrow \chi_{c2}K)$,
$\mathcal {B}(B_c\rightarrow \chi_{c2}\rho)$,  and $\mathcal {B}(B_c\rightarrow \chi_{c2}K^*)$, as $7.5\times 10^{-5}$,
$5.49\times 10^{-6}$, $2.38\times 10^{-4}$, and $1.33\times 10^{-5}$, respectively,
which are considerably smaller than our results as well as most of other model calculations.
Our results  for a final $K^{(*)}$  are also larger than those of other approaches.
The disagreement in the predictions may be attributed to the
different values of the form factors used in these approaches.
Experimental investigations on these decays may be used to test theoretical methods according to their predictions.

On the experimental side, so far only the evidence  for the decay
$B^+_c\rightarrow \chi_{c0}\pi^+$ is found at $4.0 \sigma$  significance by the LHCb Collaboration \cite{prd94091102}.
The ratio of production cross sections of the $B^+_c$ and $B^+$ mesons times branching fractions is measured to be
$\frac{\sigma_{B_c^+}}{\sigma_{B^+}}\times \mathcal {B}(B^+_c\rightarrow \chi_{c0}\pi^+)
=(9.8^{+3.4}_{-3.0}(\text{stat})\pm 0.8(\text{syst}))\times 10^{-6}$ \cite{prd94091102}.
As a cross-check,  the   cross section  ratio $\frac{\sigma_{B_c^+}}{\sigma_{B^+}}$ can be extracted from 
another charmonium mode,  $\frac{\sigma_{B_c^+}}{\sigma_{B^+}}\times
\frac{\mathcal {B}(B^+_c\rightarrow J/\psi \pi^+)}{\mathcal {B}(B^+\rightarrow J/\psi K^+)}=(0.683\pm0.018\pm0.009)\%$  measured by the
LHCb Collaboration \cite{prl114132001}. The branching ratio $\mathcal {B}(B^+\rightarrow J/\psi K^+)$, determined from the world average value,
is $(1.026\pm0.031)\times10^{-3}$ \cite{pdg2016}.
If we use our previous PQCD calculation  $\mathcal {B}(B^+_c\rightarrow J/\psi \pi^+)=(2.33^{+0.81}_{-0.61})\times 10^{-3}$ \cite{rui1},
where all errors are combined in quadrature, as an input, the ratio  $\frac{\sigma_{B_c^+}}{\sigma_{B^+}}$ is in the region of
$(2.2\sim 4.1)\times 10^{-3}$. 
Combined with the prediction on  $\mathcal {B}(B^+_c\rightarrow \chi_{c0} \pi^+)$ in Table \ref{tab:br1}, we obtain the range  $\frac{\sigma_{B_c^+}}{\sigma_{B^+}}\times \mathcal {B}(B^+_c\rightarrow \chi_{c0}\pi^+)=(2.6\sim 8.2)\times 10^{-6}$,
which is consistent with the LHCb data  with one sigma errors.

\begin{table}
\caption{The PQCD predictions for the  polarization fractions, relative phases in the
$B_c\rightarrow (A,T) V$  decays. The errors induced by the same sources as in Table \ref{tab:br1}}
\label{tab:po1}
\begin{tabular}[t]{l cccccc}
\hline\hline
Modes   & $f_0$ & $f_{\parallel}$ &$ f_{\perp}$ &$ \phi_{\parallel}(\text{rad})$&$\phi_{\perp}(\text{rad})$ \\ \hline
 $B_c^{+} \rightarrow  \chi_{c1}\rho^{+}$  & $0.66^{+0.03+0.04+0.00}_{-0.02-0.04-0.00}$ & $0.15^{+0.02+0.03+0.00}_{-0.01-0.02-0.00}$
  & $0.18^{+0.03+0.03+0.01}_{-0.01-0.02-0.00}$  &$1.21^{+0.06+0.08+0.01}_{-0.04-0.09-0.01}$ &$1.67^{+0.04+0.04+0.00}_{-0.04-0.04-0.01}$   \\
   $B_c^{+} \rightarrow  \chi_{c1}K^{*+}$  & $0.60^{+0.03+0.05+0.02}_{-0.03-0.04-0.00}$ & $0.18^{+0.02+0.02+0.00}_{-0.02-0.03-0.01}$
  & $0.22^{+0.02+0.03+0.00}_{-0.01-0.03-0.00}$  &$1.22^{+0.06+0.08+0.00}_{-0.04-0.10-0.03}$ &$1.68^{+0.04+0.04+0.01}_{-0.05-0.04-0.01}$   \\
   $B_c^{+} \rightarrow  \chi_{c2}\rho^{+}$  & $0.93^{+0.01+0.01+0.00}_{-0.02-0.02-0.01}$ & $0.05^{+0.00+0.01+0.00}_{-0.00-0.01-0.00}$
  & $0.03^{+0.00+0.00+0.00}_{-0.01-0.01-0.01}$  &$1.00^{+0.06+0.01+0.01}_{-0.04-0.02-0.02}$ &$1.12^{+0.05+0.01+0.02}_{-0.05-0.01-0.04}$   \\
   $B_c^{+} \rightarrow  \chi_{c2}K^{*+}$  & $0.90^{+0.01+0.01+0.01}_{-0.00-0.01-0.01}$ & $0.06^{+0.01+0.01+0.01}_{-0.00-0.00-0.00}$
  & $0.03^{+0.00+0.01+0.01}_{-0.00-0.00-0.00}$  &$1.00^{+0.06+0.03+0.01}_{-0.03-0.01-0.02}$ &$1.12^{+0.06+0.01+0.02}_{-0.05-0.00-0.04}$   \\
  $B_c^{+} \rightarrow  h_{c}\rho^{+}$  & $0.91^{+0.01+0.02+0.01}_{-0.01-0.02-0.01}$ & $0.04^{+0.01+0.01+0.00}_{-0.01-0.02-0.01}$
  & $0.05^{+0.01+0.01+0.01}_{-0.01-0.00-0.00}$  &$0.83^{+0.05+0.02+0.00}_{-0.05-0.03-0.01}$ &$1.11^{+0.04+0.02+0.01}_{-0.04-0.03-0.01}$   \\
   $B_c^{+} \rightarrow  h_{c}K^{*+}$  & $0.88^{+0.01+0.02+0.01}_{-0.01-0.00-0.00}$ & $0.05^{+0.01+0.00+0.00}_{-0.01-0.01-0.01}$
  & $0.07^{+0.01+0.00+0.00}_{-0.01-0.01-0.01}$  &$0.85^{+0.05+0.01+0.00}_{-0.05-0.01-0.01}$ &$1.13^{+0.04+0.02+0.01}_{-0.04-0.02-0.01}$   \\
\hline\hline
\end{tabular}
\end{table}

Turning to the polarizations 
for $B_c \rightarrow AV, TV$ decays.
 We usually define five observables corresponding to three polarization fractions $f_{\lambda}(\lambda=0,\parallel,\perp)$,
and two relative phases $\phi_{\parallel}, \phi_{\perp}$, where
 \begin{eqnarray}\label{eq:jihua}
f_{\lambda}=\frac{|\mathcal {A}_{\lambda}|^2}{|\mathcal {A}_0|^2+|\mathcal {A}_{\parallel}|^2+|\mathcal {A}_{\perp}|^2},\quad
\phi_{\parallel,\perp}=\text{arg}\frac{\mathcal {A}_{\parallel,\perp}}{\mathcal {A}_{0}},
\end{eqnarray}
with normalization such that $\sum\limits_{\lambda}f_{\lambda}=1$.
The results for the   polarization fractions
and their relative phases
are displayed in Table \ref{tab:po1}, where the sources of the errors in the
 numerical estimates have the same origin as in the discussion of the branching ratios in Table \ref{tab:br1}.
 It can be observed that both the polarization fractions and the phases 
 are relatively stable with respect to the variations of hadronic parameters, the decay constants and the hard scale,
 and therefore they serve as  good quantities to test the standard model.
Several remarks are given in order. First, the contributions to the  branching
ratios mainly arise from the longitudinal polarizations 
because of the relation $f_0\gg f_{\parallel}\sim f_{\perp}$, which is expected
from the power counting rules.
For example,
the longitudinal parts of $B_c\rightarrow TV$  decays occupy over $90\%$,
which are very similar to the case of $B_c\rightarrow J/\psi V$ \cite{rui1}.
However,  the longitudinal polarizations of $B_c\rightarrow \chi_{c1}V$ are relative smaller ($\sim60\%$)
compared to that of $B_c\rightarrow h_c V$.
As mentioned before, owing to the G-parity, the  distribution amplitudes  for $\chi_{c1}$ and
$h_c$  mesons exhibit the different asymptotic behaviors (see Eqs. (\ref{eq:3p1}) and (\ref{eq:1p1})).
If we use the $h_c$ distribution amplitudes for calculation,  the resultant predictions  $f_0(\chi_{c1}V)$
 can be increase to around $90\%$. 
 Besides, the longitudinal and transverse decay constants in the two axial-vector mesons can also contribute to
 different polarizations.
Second, for $B_c\rightarrow (A,T)\rho$ and $B_c\rightarrow (A,T)K^*$ decays,
both   have similar magnitudes and phases of the amplitudes, which
suggests the SU(3) breaking effect between them is small.
Last, the predicted relative phases deviations from $\pi$
indicate the existence of the still unknown final-state interaction.
 However, the magnitudes and phases of the two transverse
amplitudes $\mathcal {A}_{\parallel}$ and $\mathcal {A}_{\perp}$ are roughly equal,
which is expected from analyses based on quark-helicity conservation \cite{zpc1269,prd64117503}.
These results and findings will be further tested by the LHCb and
Belle-II experiments in the near future.

\section{ conclusion}\label{sec:sum}

The two-body  $B_c$ meson decays to a $P$-wave charmonium state $(\chi_{c0},\chi_{c1},\chi_{c2},h_c)$ and a light $(\pi,K,\rho,K^*)$ meson
 are systematically  analysed within the  perturbative QCD approach.
Our predictions for the branching ratios are summarized in Tables \ref{tab:br1} and \ref{tab:br2} and compared with other theoretical
results. Overall, the  predicted branching ratios from different theoretical models  have a relative big spread.
The upcoming experimental measurements of the corresponding decay rates can examine various theoretical approaches.
Based on our estimations, the dominating decay mode of the concerned processes is $B_c\rightarrow \chi_{c2}\rho$
with predicted  branching ratios of $1.6\%$, which should   be
accessible experimentally at high-luminosity hadron colliders.
We also estimate the polarization contributions in $B_c\rightarrow (\chi_{c1,c2},h_c)V$ decays.
As expected, based on the factorization assumption,
the longitudinal polarization dominates and
the transverse polarizations are of the same size.

We also discussed theoretical uncertainties arising from
the hadronic parameters in $B_c$ meson wave function, the decay constants of  charmonium  states and the hard scale $t$.
The branching ratios suffer a large error from the decay constants, whereas the polarization observables are less sensitive
to these parameters. The obtained results can be confronted to the experimental data in the future.

\begin{acknowledgments}
I would like to acknowledge Hsiang-nan Li and Yu-Ming
Wang for helpful discussions. This work is supported in
part by the National Natural Science Foundation of China
under Grants No. 11547020 and No. 11605060 and, in part,
by the Program for the Top Young Innovative Talents of
Higher Learning Institutions of Hebei Educational
Committee under Grant No. BJ2016041.
\end{acknowledgments}

\begin{appendix}
\section{DETAILS FOR DERIVING THE P-WAVE CHARMONIUM DAS}\label{sec:das1}
Starting with the  momentum-space radial wave function which can be written as  the Fourier transform of the position-space expression $\psi_{nlm}(\vec{r})$
\begin{eqnarray}\label{eq:fuliye}
\psi(k)=\int_{-\infty}^{\infty} \psi_{nlm}(\vec{r})e^{-i\vec{k}\cdot \vec{r}}d\vec{r},
\end{eqnarray}
where $n$, $l$, and $m$ stand for main, orbital, and magnetic quantum  numbers, respectively.
In above equation, the first term  $\psi_{nlm}(\vec{r})$ is known to be separated into $R_{nl}(r)Y_{lm}(\theta,\varphi)$
 in the spherical coordinates ($r$, $\theta$, $\varphi$),
while  the second exponential term  in the plane wave expansion can be written as
\begin{eqnarray}
e^{-i\vec{k}\cdot \vec{r}}=e^{-ik r \cos \theta}=\sum_{l'=0}^{\infty} \sqrt{4\pi(2l'+1)}(-i)^{l'}j_{l'}(kr)Y_{l'0}(\theta,0),
\end{eqnarray}
with $j_{l'}(kr)$  the spherical Bessel function.
We then write  Eq.(\ref{eq:fuliye})  as
\begin{eqnarray}\label{eq:fuliye1}
\psi(k)= \sqrt{4\pi(2l+1)}(-i)^{l}  \int_{0}^{\infty}j_{l}(kr) R_{nl}(r) r^2 dr,
\end{eqnarray}
where  the orthogonality property 
$\int_0^{\pi} \int_0^{2\pi}Y_{lm}Y_{l'0}\sin \theta d\theta d\varphi=\delta_{ll'}\delta_{m0}$
have been used.

For the $P$-wave states $ n=2$ and $l=1$, employing the spherical Bessel function $j_1(kr)=\frac{\sin(kr)-kr\cos(kr)}{(kr)^2}$
and the radial wave function  for a Coulomb Potential $R_{21}(r)\propto r e^{\frac{-q_B r}{2}}$,
 the integral of Eq. (\ref{eq:fuliye1}) evaluates to
\begin{eqnarray}\label{eq:fuliye2}
\psi(k) \propto \frac{k q_B}{(4k^2+q_B^2)^3},
\end{eqnarray}
where $q_B$ is the Bohr momentum.
Note that the above expression  is in contrast to   Eq.(47) in \cite{prd71114008}.
We argue that  the spherical harmonics function for $P$-wave states is dependent on the angle $\theta$, which should  contribute to the
integral in Eq. (\ref{eq:fuliye}).
In particular,  Eq.(\ref{eq:fuliye2}) is almost the same as M. Beneke's calculation in Ref. \cite{npb811155} (see Eq.(45)),
except for a constant term which can be absorbed in the redefinition of the wave function of the $P$-wave charmonium.
Following much the same procedure as described in Refs \cite{plb612215,prd71114008},
we obtain the heavy quarkonium DA which is dependent on the charm quark momentum fraction $x$ after integrating the transverse
momentum $k_T$,
\begin{eqnarray}\label{eq:fuliye3}
\Phi(x)\sim \int d^2k_T \psi(x,k_T)\propto x(1-x)\{  \frac{\sqrt{x(1-x)(1-4x(1-x))^3}}{[1-4x(1-x)(1-v^2/4)]^2}\},
\end{eqnarray}
where $v=q_B/m_c$ is the charm quark velocity. In the numerical calculation, we take $v^2=0.3$ and neglect the $v^2$ term
 in the numerator  \cite{prd71114008}.
 As mentioned in Eq. (\ref{eq:asym}), we propose the $P$-wave charmonium states DAs  as
$\psi(x)\propto \Phi_{asy}(x)\mathcal {T}(x)$
with
 \begin{eqnarray}\label{eq:vv2}
\mathcal {T}(x)=\{\frac{\sqrt{x(1-x)(1-4x(1-x))^3}}{[1-4x(1-x)(1-v^2/4)]^2}\}^{1-v^2},
\end{eqnarray}
where the power $1-v^2$ denotes the small relativistic corrections to the Coulomb wave functions \cite{plb612215}.

\end{appendix}

\end{document}